\documentclass[prb,twocolumn,amsmath,amssymb,amsfonts,superscriptaddress,floatfix,showpacs]{revtex4-1}
\usepackage{graphicx}
\usepackage{bm}
\usepackage{amsmath}
\begin{document}
\title{A first principles study of magnetism in Pd$_{3}$Fe under pressure}
 
\author{Biswanath Dutta} 
\affiliation{Department of Physics, Indian Institute of Technology
Guwahati, Guwahati, Assam 781039, India } 
\author{Sumanta Bhandary}
\affiliation{Department of Physics and Astronomy, Uppsala University, Box-516,
SE 75120, Uppsala, Sweden}
\author{Subhradip Ghosh}
\affiliation{Department of Physics, Indian Institute of Technology
Guwahati, Guwahati, Assam 781039, India } 
\author{Biplab Sanyal} 
\email{For correspondence: Biplab.Sanyal@physics.uu.se}
\affiliation{Department of Physics and Astronomy, Uppsala University, Box-516,
SE 75120, Uppsala, Sweden}

\date{\today}

\begin{abstract}
Recent experiments on Pd$_{3}$Fe intermetallics [Phys. Rev. Lett.
102, 237202 (2009)] have revealed that the system
behaves like a classical invar alloy under high pressure. The experimental
pressure-volume relation suggests an anomalous volume collapse 
and a substantial increase in bulk modulus around the pressure
where invar behavior is observed. With the help of first-principles density functional
theory based calculations, we have explored various magnetic phases (ferromagnetic, fully and partially disordered local moment, spin spiral) in order to
understand the effect of pressure on magnetism. Our calculations reveal that the
system does not undergo a transition from a ferromagnetic to a spin-disordered state
as was thought to be the possible mechanism to explain the invar behavior of this system.
We rather suggest that the anomaly in the system could possibly be
due to the transition from a collinear state to  non-collinear magnetic states
upon the application of pressure. 
\end{abstract}

\maketitle

\section{Introduction}
The technological usefulness of the state-of-the-art magnetic materials depends on their
properties at finite temperature and pressure. Fe-based alloys and intermetallics 
have their rich history of being used as smart magnetic materials \cite{mag}. Numerous
experimental and theoretical investigations have been carried out to understand the
magnetic properties of pure Fe and its alloys and intermetallics at finite temperatures \cite{mag1,mag2,mag3}(and references therein). A very special characteristic of some Fe-based alloys is observed as the invar properties where the thermal expansion coefficient remains invariant with temperature. A wealth of literature on theoretical and experimental results on invar alloys exist. \cite{mag1, mag2, mag3, igornature, invar, invar1, invar2, invar3, sanyal} Besides the conventional temperature invariant invar properties, pressure induced invar characteristics have also been reported. \cite{noncolin} 

In this paper, we will discuss an Fe-based alloy where pressure induced invar properties have been observed in experiments recently. \cite{prl} EDXAD experiments on Pd$_{3}$Fe alloys have 
reported an anomalous volume collapse under pressure between
10 and 15 GPa \cite{prl}. The EDXAD data were successfully fit to a equation of state which
points towards the existence of two separate states; one at high pressure and the other
at low pressure. Subsequent measurements at high temperatures reveal that at 7 GPa
pressure, there is an anomalously low thermal expansion suggesting that the system behaves
like an Invar alloy \cite{invar} at high pressure. First principles calculations to understand
the origin of these anomalies were rather inconclusive. However, in order to explain
the invar behavior under pressure in this system, it was conjectured that the 
application of pressure may reduce the Curie temperature below room temperature, thus
making possible a transition in magnetic state from ferromagnetic to a spin-disordered
one and that the invar anomaly is a result of such high volume-high spin to 
low volume-low spin transition \cite{prl}. Such a proposition stems from the 
facts that in Fe-rich
FePt and FePd alloys, the classical invar anomaly at ambient pressure was explained
in terms of the transition from a high volume ferromagnetic state to a low volume
spin disordered state \cite{invar1,invar2,invar3}. In case of Fe-based alloys 
with low Fe concentration, the
magnetically ordered state was found to be stabilized and thermal invar behavior
was suppressed, in general. However, it was discovered that in case of Fe-Ni alloys with
low Fe concentration, the alloy with normal thermal expansion properties at ambient
pressure exhibits invar behavior at high pressure \cite{noncolin}. 
This explanation for this anomaly
was given in terms of transition to non-collinear configurations at lower volumes
due to application of pressure. On the other hand, Khmelevsky {\it et al.}
\cite{rsg} investigated the magnetic transitions in Fe$_{0.7}$Pt$_{0.3}$
under pressure and ruled out the possibility of making a connection
between the observed invar anomaly and a transition to noncollinear states
from zero pressure collinear one.
These results, thus, indicate that the physical origin of anomalies
related to the magnetic properties in Fe-based alloys under pressure may
not be unique.
In this paper, we, therefore, aim at understanding the magnetism in Pd$_{3}$Fe alloys
under pressure by extensive quantitative analysis of the energetics, the magnetization,
the pressure-volume relations with different magnetic structures. To this end
we have performed first principles calculations with the collinear 
magnetic structures viz. ferromagnetic, antiferromagnetic,
and the the spin disordered structures and with the spiral
magnetic structures. Our results show
that contrary to the proposition in Ref.~\onlinecite{prl}, the volume collapse and the subsequent
invar anomaly under pressure, observed experimentally, cannot be explained by a
transition from magnetically ordered state to a spin disordered state. Rather, the
results indicate that the anomaly could be due to the transition from a collinear
magnetic state to noncollinear ones.

The paper is organized as follows. In Section II, we briefly discuss the details of
computational methods used. In Section III, we present our results on various magnetic
structures. A detailed discussion followed by the conclusions are presented at the
end. 
\section {Computational Details}
All calculations have been performed within the
standard framework of the spin polarized version of the density 
functional theory \cite{kohn,kohn1}.
The spin disordered magnetic structure has been simulated using the disordered local
moment (DLM) formalism \cite{dlm} where the magnetic disorder is represented by considering
the Pd$_{3}$Fe system as a pseudoternary alloy Pd$_{3}$Fe$^{+}_{x}$Fe$^{-}_{1-x}$,
where $x$ is the concentration of Fe atoms with up spin (Fe$^{+}$) and $1-x$ is 
with down spin (Fe$^{-}$). The paramagnetic or the full DLM state,
henceforth mentioned as FDLM, is obtained when $x=0.5$. Partial disordered
local moment (PDLM) is realized for $x \ne 0.5$. 
Calculations for the DLM states are performed by the
Green's function based Exact Muffin-tin Orbital (EMTO) \cite{emto1,emto2,emto3} formalism in conjunction with
the Coherent Potential Approximation (CPA) \cite{Taylor}, a single site mean field formalism to
treat the chemical and magnetic disorder. 
The EMTO-CPA is a suitable method for the treatment of the
spin disordered state as the usage of conventional electronic structure methods
require construction of prohibitively large supercell to simulate the magnetically
disordered states. The other advantage of the EMTO method is that it uses two different
approaches for calculations of the one-electron states and the potentials:while the
one electron states are calculated exactly for the overlapping muffin-tin potentials,
the solution of the Poisson's equation can include certain shape approximations.
An accuracy at a level comparable to the full-potential techniques can, thus, be
sustained, without a significant loss of accuracy. However, since the present
implementation of the EMTO method does not address noncollinear magnetic ordering,
the spiral magnetic structures have been studied by the full-potential linearized
augmented plane wave (FP-LAPW) method \cite{lapw1} using the ELK code \cite{elk} in an implementation 
which allows noncollinear magnetism 
including spin spirals \cite{lapw2,lapw3}. 
In this implementation, the full magnetization density is
used in addition to the full charge density and the full potential. The magnetic
moment is allowed to vary both in magnitude and in direction inside the atomic spheres
as well as in the interstitial regions.

All calculations have been performed within the 
local spin density approximation (LSDA)\cite{lsda}
for the exchange-correlation part of the Hamiltonian. For the EMTO-CPA calculations,
the Green's functions were calculated for 32 complex energy points distributed
exponentially on a semi-circular contour. The total energies were calculated by the
full charge density technique \cite{fcd}. 
The $s, p$ and $d$ orbitals were included in the EMTO
basis set for the expansion of the wave functions. The Brillouin zone integrations were
performed with 1540 {\bf k}-points. In the FP-LAPW method, the muffin tin radii of
Fe and Pd were taken to be 1.16 and 1.27 \AA~respectively. The plane wave
cutoff for basis functions is $RK_{max}=8$.
The cutoff for the charge density was considered to be $G_{max}$=12. The number
of {\bf k}-points for the Brillouin zone integrations were 364. A broadening of
0.14 eV was used according to Methfessel-Paxton \cite{mp}scheme. 
Total energies were converged to less than 1 meV per atom for all the
calculations.
\section{Results and Discussions}
In Ref.~\onlinecite{prl}, pressure-volume relations for Pd$_{3}$Fe showed an anomalous
reduction in volume from $V_{red}=\frac{V}{V_{0}}=0.96$ to $0.91$ 
between 10 and 15 GPa pressures, $V_{0}$ being the equilibrium
volume of the ferromagnetic state. Since the pressure-volume curve could
be smoothly fitted to a Weiss like equation of state, it was proposed
that the anomaly is related to the transition from a high volume-high spin
state to a low volume-low spin state. On the other hand, the low thermal
expansion under pressure suggested an invar anomaly 
and its connection to the two state model of Weiss \cite{weiss}.
The authors, however, could not find out a stable low spin state at lower
volume from first-principles calculations. Therefore it was proposed that the
required low volume-low spin state could be a paramagnetic state. Such
a transition, according to them, could be because of lowering of the
Curie temperature $T_{c}$ due to the application of pressure. In this section,
we present results of extensive first-principles calculations to resolve
these issues.
\begin{figure}[t]
\includegraphics[scale=0.33]{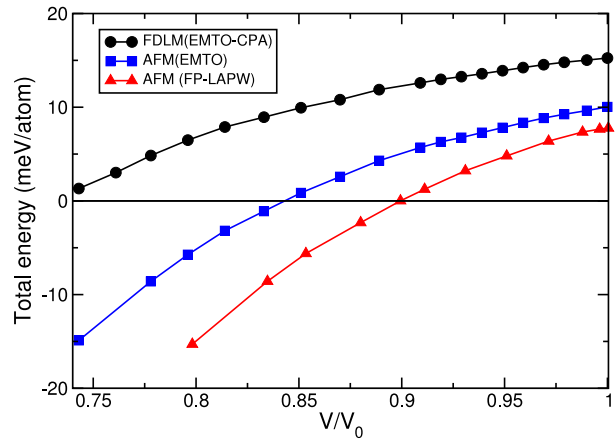}
\caption{(Color online) Total energy as a function of $V_{red}$ for FM, AFM and DLM states
computed by the EMTO-CPA method. The energy of the ferromagnetic state is
taken to be zero. The results of FP-LAPW calculations for FM and AFM states
are shown for comparison.}
\label{fig1}
\end{figure}

\begin{figure}[t]
\includegraphics[scale=0.33]{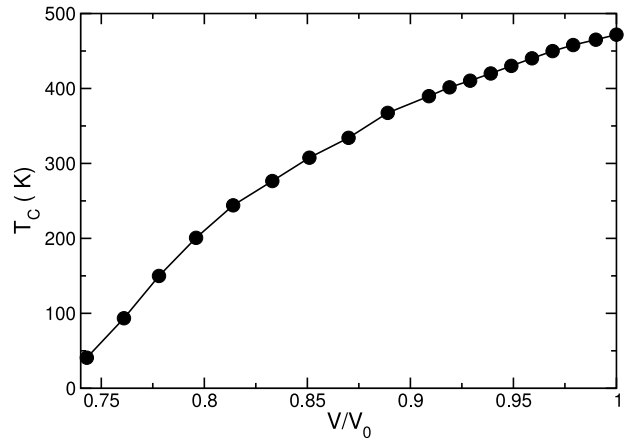}
\caption{Curie temperature $T_{c}$ as a function of $V_{red}$ computed
by the EMTO-CPA method.}
\label{fig2}
\end{figure}

\begin{figure}[t]
\includegraphics[scale=0.35]{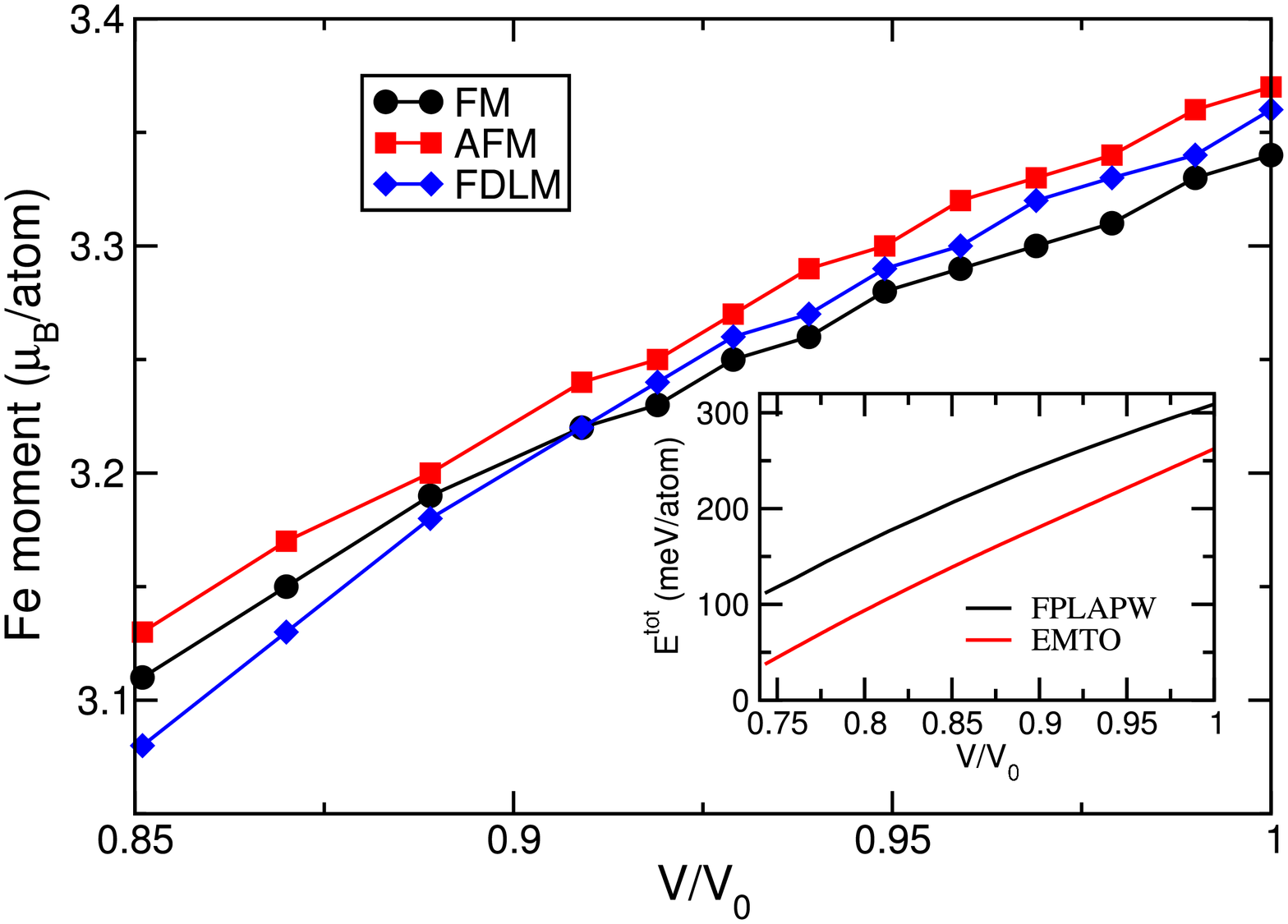}
\caption{(Color online) Local moments of Fe for FM, AFM and FDLM states as a function
of compression computed by the EMTO-CPA method. (Inset) Total energy of LS state in meV per atom with respect to FM state as a function of compression computed by the FP-LAPW and the EMTO methods.}
\label{fig3}
\end{figure}

\begin{figure}[t]
\includegraphics[scale=0.34]{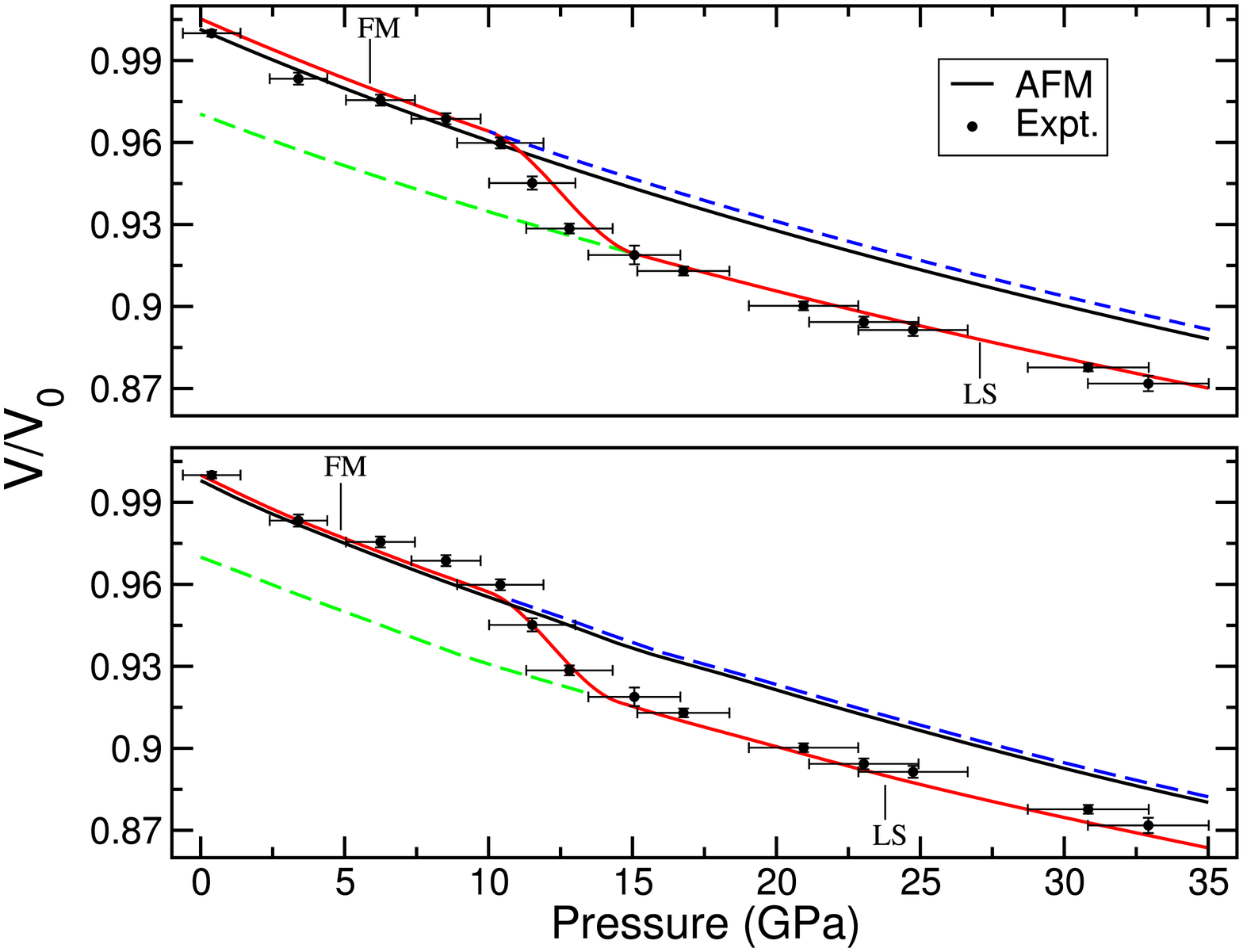}
\caption{(Color online) Pressure-volume relations for Pd$_{3}$Fe computed by the
FP-LAPW method (top panel) and the EMTO method (bottom panel). Results
of FM (blue dashed), AFM (black) and LS (green dashed) states are shown in the plot.
Experimental results (red line with error bars) are included for comparison.}
\label{fig4}
\end{figure}

In Fig.~\ref{fig1},
we present results on the relative stabilities of ferromagnetic (FM),
antiferromagnetic (AFM) and paramagnetic (PM) states by EMTO, EMTO-CPA
and FP-LAPW methods. PM state has been modeled by a full disordered local moment (FDLM)
treatment. The AFM state, considered for comparison, has Fe moments
ordering ferromagnetically in alternating (110) planes. We have done
calculations for two other AFM states, in one of which the Fe moments order
ferromagnetically in alternating (100) planes and in the other in
alternating (111) planes. The energy of the first one, the (110) AFM was
the lowest and hence it has been included for comparison. The results of the
EMTO and EMTO-CPA calculations show that the AFM state becomes stable in
comparison to the FM state at $V_{red} \sim 0.85$, while the DLM state
would be stable at a much higher compression. The FP-LAPW calculations
predict that the AFM state becomes stable at $V_{red}=0.9$. It is to be
noted that the first-principles projector augmented wave calculations of
Ref.~\onlinecite{prl} also observed stabilization of the AFM state at around $V_{red}=0.9$.
The proximity of the FP-LAPW and the projector augmented wave results
are not surprising since both solve the Kohn-Sham equations without
invoking any shape approximations on the charge density and potential.
A different  $V_{red}$ value for FM-AFM transition as computed
by the EMTO is due to the fact that certain degree of shape approximations
is inherent in the muffin-tin orbital based methods. However, this
discrepancy does not seriously affect the physics as our primary concern
here is to find out a magnetic phase, energetically lower than the FM state
at around the experimental pressure where the anomalous behavior in
pressure-volume relation is found and relate the stability of the phase
with the anomalous behavior. It is to be noted that the $V_{red}$ values
of 0.9 obtained by the FP-LAPW or of 0.85 obtained by the EMTO
for the FM-AFM transition are quite close to the $V_{red}$ value of 0.91
where the anomalous volume collapse was observed in EDXD measurements. 
However, the outcome that the FDLM state is
having a energy higher than the FM state until a much larger compression,
rules out the proposition of Ref.~\onlinecite{prl}, that the FM state
transforms to the paramagnetic state upon compression with a lowering of
$T_{c}$. In order to validate our argument, we have calculated the
$T_{c}$ as a function of the compression, as shown in Fig.~\ref{fig2}. 
The $T_{c}$ has been estimated by the mean-field approximation as,
\begin{eqnarray}
k_{B}T_{c} & = & \frac{2}{3}\left(E_{FDLM}-E_{FM}\right)
\end{eqnarray}
where $E_{FDLM},E_{FM}$ are the total energies of the system in FDLM state
and in the ferromagnetic states respectively. The calculated $T_{c}$ of 472K compares well with the
experimental value of 499K \cite{tc} at zero pressure ($V_{red}=1$). The results
suggest that the $T_{c}$ indeed decreases with pressure and reduces
below room temperature at around $V_{red}=0.85$. This, though, may tempt
one to infer that at this compression, the system should undergo a transition from
FM to PM state, making a correlation between this result and
the anomalous volume collapse, observed experimentally, requires two things:
first, the zero pressure volume of the PM state
should be significantly lower than the FM state, introducing volume
magnetostriction as is observed in the FePt and FePd alloys with high Fe
concentrations where the invar anomaly is observed \cite{invar1} and 
second,the
bulk modulus of the low volume state should be substantially higher than
the high volume FM state as was observed experimentally \cite{prl}. In Table I,
we present results on zero pressure lattice constant and bulk modulus
computed for the FM, the AFM and the DLM states. Results obtained by the
EMTO and the FP-LAPW methods along with the theoretical results of Ref.~\onlinecite{prl}
are presented for comparison.
The results clearly show that the equilibrium lattice constants and
the bulk modulus of the
FM, AFM and FDLM states calculated by the EMTO-CPA method are nearly
equal, thus leaving no scope of observing magnetostriction and the
features in bulk modulus observed in experimental data of Ref.~\onlinecite{prl}. The
FP-LAPW calculations for FM and AFM states also imply the same trend
with regard to these two magnetic states. Another feature of the invar
anomaly observed in case of Fe$_{3}$Pt alloy was that the low volume
FDLM state had significant reduction in the magnetic moment at the Fe
site in comparison to that in the FM state \cite{invar1}. 
The magnetovolume effect
connected to this reduction of Fe moment in the FDLM state was the
reason behind the invar anomaly \cite{invar1,invar3}. 
To explore this aspect, in Fig.~\ref{fig3},
we show the variation of the magnetic moment at the Fe site with 
compression for the FM, the AFM and the FDLM states. The results suggest
that the Fe moments are extremely localized and there is hardly any
deviation of the Fe local moment across the magnetic orders. All these
observations, particularly regarding the behavior of structural and
magnetic properties in the FM and the FDLM states, suggest that
the experimentally observed features under pressure in Pd$_{3}$Fe are
not due to a transition from the FM to a paramagnetic(FDLM) state. Also, we show
a comparison between the relative total energies obtained in EMTO and FPLAPW
methods in the inset of Fig.~\ref{fig3}. In both methods, FM state has energy lower than that of LS
state and the trend is similar as a function of compression.
In Table~I, we have also presented results on the equilibrium lattice
constant and the bulk modulus for a low-spin(LS) state. The calculations
for this state was done by keeping the spins collinear but by fixing the
total magnetic moment as $0.01 \mu_{B}$. The consideration of such a state
was part of the attempt to search for a low volume low spin state which
may be related to the explanation of the anomalies observed experimentally.
The results suggest that the equilibrium volume of this LS state is
significantly lower than the FM state and the bulk modulus considerably
higher. In Fig.~\ref{fig4}, we plot the pressure-volume relations computed
by both EMTO and FP-LAPW methods. The results show a striking
agreement between the experiment and the theory, both qualitatively, as well
as quantitatively when one considers only the FM and the LS states. 
The pressure-volume relation for AFM state is also
shown for comparison. It appears that the desired anomalous reduction in
volume between 10-15 GPa can be explained by a transition from FM to LS
state. However, upon calculating the total energy of the LS state we find
that the LS state is not stable in comparison to the FM state in the desired
pressure regime as shown in Fig.~\ref{fig4}. In fact, the energy of the LS state
is more than 100 meV /atom higher than the FM state around the compression
(inset of Fig.~\ref{fig3})
where the anomaly is observed and that it becomes stable only at a much greater
compression than the FDLM state. Fig.~\ref{fig4} suggests that this qualitative
feature is independent of the method of calculation used. This is also
reported in Ref.~\onlinecite{prl}. Thus the excellent agreement between the theory and
the experimental pressure-volume relations can be purely fortuitous and can
not be used as a viable explanation of the anomalies. 

\begin{table}
\caption{The equilibrium(zero pressure) lattice constants $a_{0}$ in \AA~and 
bulk modulus $B_{0}$ in GPa for different magnetic states computed by different
electronic structure methods.} 
\begin{ruledtabular} 
\begin{tabular}{lcccccc}
 Magnetic &~~~EMTO-CPA&~~~~~FP-LAPW&~~~~Ref.~\onlinecite{prl} \\ 
 state &$a_{0}$~~~~~~~~~~~$B_{0}$&~~~~$a_{0}$~~~~~~~~~$B_{0}$&~~~~$a_{0}$~~~~~~$B_{0}$\\ \hline
 FM &3.888~~~~~~~~204.8&~~~~3.794~~~~~~216.7&~~~~3.802~~~216.8  \\
 AFM &3.886~~~~~~~~202.6&~~~~3.792~~~~~~218.8&~~~~3.798~~~216.4  \\
 LS &3.850~~~~~~~~229.0&~~~~3.756~~~~~~242.3&~~~~3.760~~~242.6  \\
 FDLM &3.887~~~~~~~~206.1&~~~~----~~~~~~---- &~~~~----~~~~~----  \\
\end{tabular}
\end{ruledtabular}
\end{table}

In order to resolve the observed discrepancy between the
theoretical and experimental results, we next consider the following
two possibilities: (i) upon application of pressure, the system does
not make a transition from FM to a FDLM state but to a partially spin
disordered state (PDLM) and (ii) upon application of pressure, the
system may not retain the perfect chemical order, instead the system may tend
to be partially ordered or fully disordered. 

\begin{figure}[h]
\includegraphics[scale=0.33]{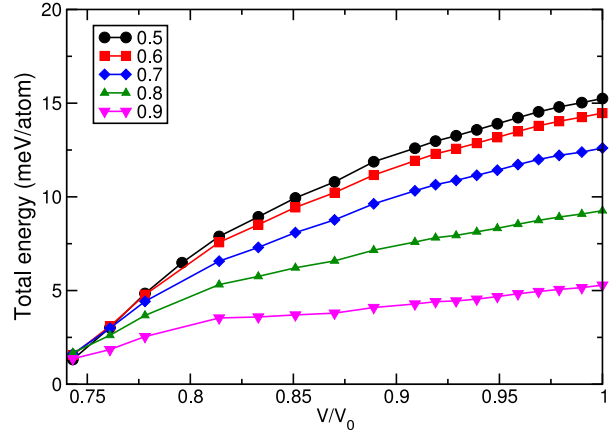}
\caption{(Color online) Variation of the total energies for various PDLM states as a
function of compression for fully ordered ($S=1$) Pd$_{3}$Fe. Different
PDLM states are marked by the corresponding $n_{red}$ values. The energies are measured
relative to the energy of the FM configuration.}
\label{fig5}
\end{figure}

To check the first
possibility, we have calculated the total energies of the various PDLM
states with $n_{red}=\frac{n^{Fe}_{+}}{n^{Fe}_{total}}$ being equal to
0.6, 0.7, 0.8 and 0.9 where $n^{Fe}_{+}$ and $n^{Fe}_{total}$ are the 
number of spin-up Fe atoms and the total number of Fe atoms ($n^{Fe}_{+}$+$n^{Fe}_{-}$) 
respectively, $n^{Fe}_{-}$ being the number of spin-down Fe atoms. 
The FDLM state is represented by $n_{red}=0.5$. In Fig.~\ref{fig5},
we present the total energies of these spin disordered states relative
to the total energy of the FM state as a function of compression. The
results suggest that none of the PDLM states is energetically lower than
the FM state for any value of volume compression. Interestingly, one may observe that 
 for $V_{red} \leq 0.75$, the difference in energies
between the PDLM states with different $n_{red}$ values are very close in energy. As clear from Fig.~\ref{fig1}, this is the volume region, where EMTO-CPA should yield a FM-FDLM transition. This suggests that the transition may not occur through a FDLM state but a combination of PDLM states with different degrees of spin-disorder. 

\begin{figure}[h]
\includegraphics[scale=0.33]{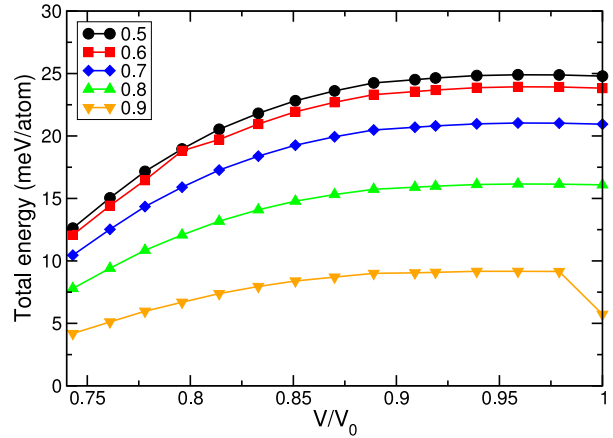}
\caption{(Color online) Same as in Fig.~\ref{fig5} but for $S=0.5$, characterizing a 
partially chemical disordered state.}
\label{fig6}
\end{figure}

\begin{figure}[h]
\includegraphics[scale=0.33]{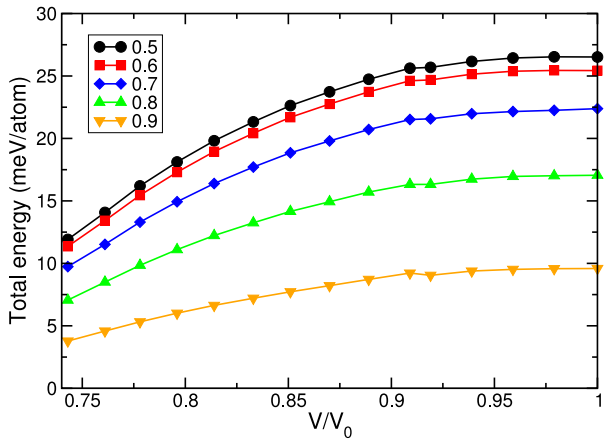}
\caption{(Color online) Same as in Fig.~\ref{fig5} but for $S=0$, characterizing a 
completely disordered state.}
\label{fig7}
\end{figure}

The second possibility as described above is explored by calculating
total energies for partial chemical order. The fully ordered
Pd$_{3}$Fe has a Cu$_{3}$Au structure with Fe on Au(I) sites and
Pd on Cu(II) sites. In the partially ordered states, Pd antisites are
introduced in the Fe (I) sites and the remaining Fe atoms are distributed
on Pd (II) sites. In order to describe the degree of chemical order, the
renormalized chemical long-range-order (LRO) parameter is defined in the
usual way for site I as
\begin{eqnarray}
S & = & \frac{c(Pd)-c^{I}(Pd)}{c(Pd)},
\end{eqnarray}
where $c(Pd)$ is total concentration of Pd in the alloy and $c^{I}(Pd)$
is a concentration of Pd atoms on Fe site I. Thus, $S=0$ corresponds to
the fully disordered alloy and $S=1$ corresponds to the fully ordered one.
In Figs.~\ref{fig6} and \ref{fig7}, we show the results of total energy calculations for
FM and various DLM states for $S=0.5$ and $S=0$ respectively. The
first noteworthy outcome of these calculations is that the equilibrium
lattice constants do not change significantly with the degree of chemical
order. The lattice constants for $S=0.5$ is 3.891~\AA~and that for
$S=0$ is 3.892~\AA~and thus hardly differ with that for $S=1$ given
in Table~I. The equilibrium bulk modulus for $S=0.5$ is 203.9 GPa and
that for $S=0$ is 199.5 GPa, once again showing a non-significant
deviation from the value obtained for $S=1$(Table I). We observe
the same trends for various DLM states as well. As was observed in case
of $S=1$, no signature of volume magneto-striction is observed for
the partially disordered and fully disordered states. The results shown
in Figs.~\ref{fig6} and \ref{fig7} suggest that same trends in relative stabilities of
PDLM and FM states as observed in Fig.~\ref{fig5} are preserved. With increasing
degree of disorder, the energy differences between $n_{red}=0.8$ and
$n_{red}=0.9$ PDLM states for the compression range $V_{red}=0.9-0.85$,
in fact increase in comparison to the fully ordered state, suggesting
that the possibility of realizing a mixed state comprising of different PDLM
states in this compression range decreases. In summary, the presence of both chemical disorder
and various degrees of spin disorder is not likely when one approaches the volume where the FM-FDLM transition takes place.

These results, therefore, indicate that the degree of chemical order and the
degree of spin disorder do not provide us any indication towards a
transition from high-volume high-spin state to a low-volume low-spin state
in Pd$_{3}$Fe under pressure. However, the results
presented so far were based upon collinear magnetic configurations. It has
been argued that the invar anomaly in Fe$_{0.64}$Ni$_{0.36}$ alloys is
related to the transition from collinear high spin state to 
noncollinear ones \cite{igornature}.
It is to be noted that the invar anomaly
accompanied by the anomaly in the pressure-volume relation and in the variation
of bulk modulus has been experimentally observed for both ambient and high pressures 
for various compositions, including compositions
with low Fe concentrations \cite{noncolin}. 
\begin{figure}[h]
\includegraphics[scale=0.33]{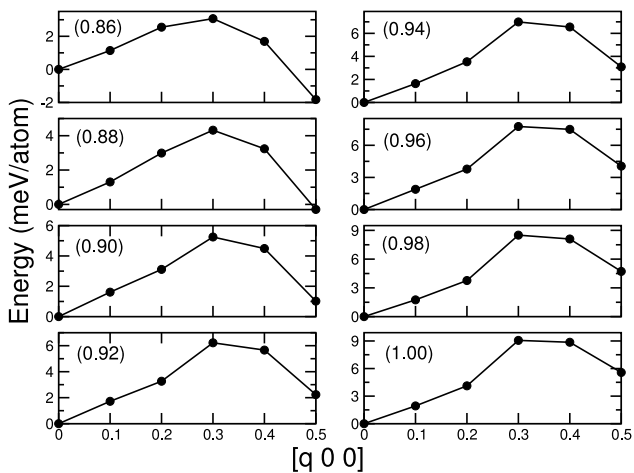}
\caption{Total energy as a function of spiral vector ${\bf q}$ in units
of $2\pi/a$ along $[\bf{q} 0 0 ]$ direction for different compressions. In the parentheses,
the values of $V_{red}$ are shown.}
\label{fig8}
\end{figure}
\begin{figure}[h]
\includegraphics[scale=0.33]{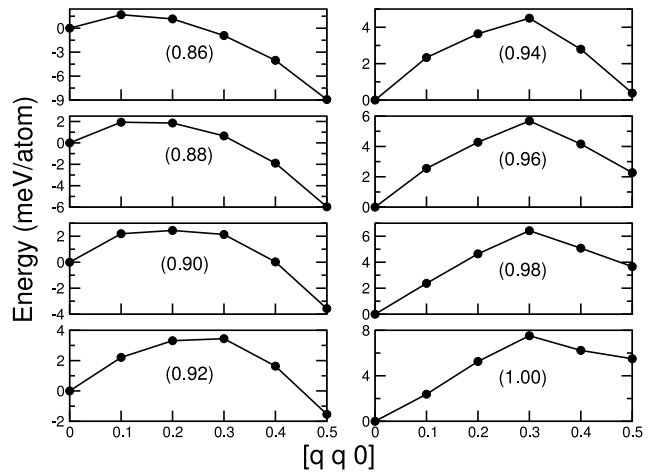}
\caption{Same as in Fig.~\ref{fig8} but for ${\bf q}$ along
$[\bf{q} \bf{q} 0]$ direction.}
\label{fig9}
\end{figure}
\begin{figure}[h]
\includegraphics[scale=0.33]{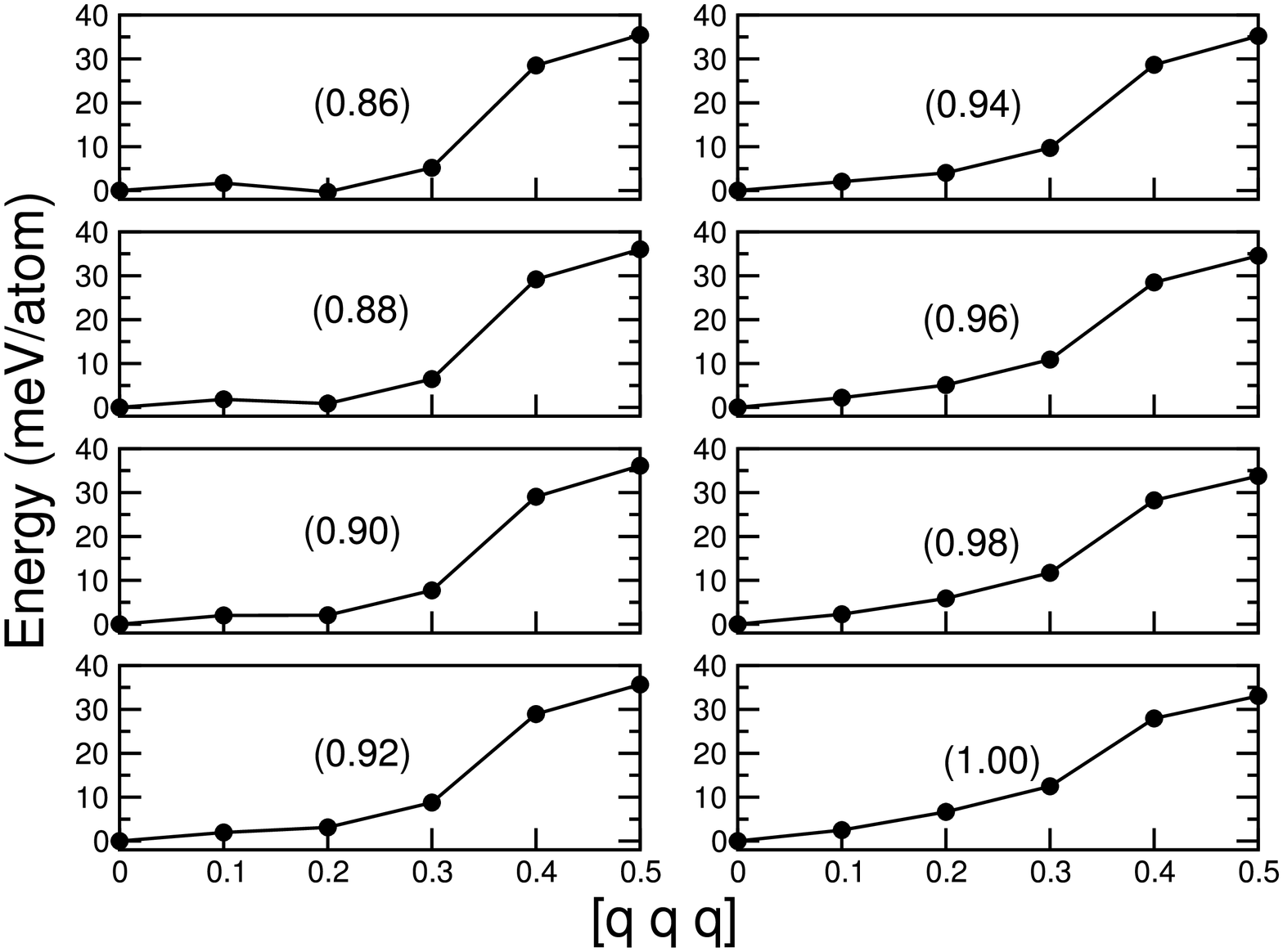}
\caption{Same as in Fig.~\ref{fig8} but for ${\bf q}$ along
$[\bf{q} \bf{q} \bf{q}]$ direction.}
\label{fig10}
\end{figure}

Following this idea, we study
the spiral magnetic states in Pd$_{3}$Fe.
Specifically, we look for the possibility of noncollinear ordering by
studying the energetics of spiral configurations. The total energy of
the system is calculated as a function of the spiral wave vector ${\bf q}$
for different values of the compression $V_{red}$. In Figs.~8, 9 and 10,
we present the results for the energetics along the high-symmetry
directions [100], [110] and [111] respectively for different compressions. 
Polesya {\it et al.} \cite{ebert} have reported the spin spiral calculations
for both ordered and disordered Pd$_{3}$Fe under ambient conditions.
Our calculated results along [100] direction for $V_{red}=1.00$ are very similar to 
their data. The behavior of the spin spiral curve is explained in terms of the induced
moment on Pd sites. Also, the reduction of Pd moment along the [100] direction
(data not shown here) is similar in both studies.

The analysis of the [100]
and [110] spin spirals show that along the [110] direction, there is a 
tendency towards the stabilization of the [0.5 0.5 0] spin spiral with
respect to the [000] one starting from $V_{red}=0.96$, making it lower
in energy than [000] one at $V_{red}=0.92$. This is consistent with the
results of the collinear calculations shown in Fig.~\ref{fig1} where the AFM
state becomes energetically stable than the FM around $V_{red}=0.9$.
Although, no incommensurate spin wave was found to be the lowest in energy,
the energy of the [0.4 0 0] spin spiral also started to decrease 
significantly from $V_{red}=0.96$, finally producing a total energy lower
than the FM phase(${\bf q}$=0 0 0) for $V_{red}=0.9$. Further compression
resulted in the complete stabilization of various incommensurate phases 
over the FM phase. Analysis for [$q$ 0 0] spin spirals suggest that although
no incommensurate phase or the AFM phase ($q$=0.5) is lower in energy
than the FM phase ($q$=0) upto $V_{red}=0.92$, the energy of the spirals
for $q \ge 0.4$ starts decreasing continuously and at $V_{red}=0.9$, they
are within 4 meV per atom in energy to the FM state. For [111] direction,
we do not see any sign of stabilization of any incommensurate or AFM phase
(${\bf q}$=0.5 0.5 0.5) with respect to FM phase. However, the energies
for $q \le 0.2$, are less than 5 meV per atom with respect to that in the
FM phase. 
\par The following picture emerges from the assimilation of all the results on
collinear magnetic states and spin spirals: first, the anomaly observed
in Pd$_{3}$Fe under pressure is not a simple effect of transition from
high volume FM state to a low volume paramagnetic or any low spin collinear
state, and second, for the range of compression, where the anomaly is seen,
there is a close competition between the FM, the AFM and other incommensurate
states. This can be interpreted from the results presented in Fig.~\ref{fig1} and
Figs.~8-10. This close competition points to the fact that the tendency
of the system is to drift from a collinear high spin state to 
increasingly noncollinear states as a function of increasing pressure. 
The anomaly in the pressure-volume 
relation and the anomaly in the bulk modulus for different compressions,
observed experimentally, act as a confirmation to this interpretation. This similar anomaly
observed in Fe$_{0.65}$Ni$_{0.35}$ at ambient
pressure \cite{igornature}and for Fe$_{0.20}$Ni$_{0.80}$ at finite pressures 
\cite{noncolin} was explained by the transition from FM to noncollinear states. 

\section{Conclusions}
We have performed extensive first-principles calculations on different
magnetic phases of Pd$_{3}$Fe under external pressure to understand the invar related pressure-volume anomaly in a recent experiment on Pd$_{3}$Fe. The results are
analyzed on the basis of relative stability of various magnetic phases as
a function of compression. Contrary to the conjecture made in Ref.~\onlinecite{prl}, we find that the experimentally observed invar anomaly in the pressure-volume
relations and bulk modulus, can not be explained based upon either the
destabilization of the equilibrium FM phase to a collinear low-spin phase
or to a paramagnetic or partially spin disordered state under application
of pressure. Non-collinear calculations with spin
spirals also could not resolve the issue by
stabilization of any incommensurate excitation under compression. However,
our results suggest that in the pressure range where the anomaly is observed in
 experiments, the FM, AFM and some incommensurate magnetic phases
compete closely, indicating that the collinear magnetic structure may be
destabilized under external pressure, making the transition to 
non-collinear structures by the anomalous reduction of volume and increase in
bulk modulus as has been observed earlier in other systems. A final
resolution on this issue can be made by first-principles calculations on
complete non-collinear magnetic structures as a function of pressure. This
will be addressed in a future communication. 

\begin{acknowledgments}
BS acknowledges G\"{o}ran Gustafssons Stiftelse and Swedish Research Council (VR) for financial support and Swedish National Infrastructure for Computing (SNIC) for granting computer time. BD acknowledges CSIR, India for financial support under the Grant-F. No. 09/731 (0049)/2007-EMR-I. BS is thankful to Lars Nordstr\"{o}m for useful discussions.
\end{acknowledgments}


\begin{thebibliography}{99}
\bibitem{mag} {\it Ferromagnetic Materials}, edited by
P. Wohlfarth and K. H. J. Buschow (North-Holland, Amsterdam, 1980-1993),
Vols. 1-7.
\bibitem{mag1} E. F. Wassermann, in {\it Ferromagnetic Materials},
edited by P. Wohlfarth and K. H. J. Buschow (North-Holland,
Amsterdam, 1990), Vol. 6.
\bibitem{mag2} M. Shiga, in {\it Materials Science and Technology},
edited by R. W. Cahn, P. Haasen and E. J. Kramer (VCH, Winheim, 1994),
Vol. 3B, Chap. 10, p. 159.
\bibitem{mag3} E. F. Wassermann and M. Acet, in {\it Magnetism and
structure in Functional Materials}, Springer series in Materials
Science Vol. 79, edited by A. Planes, L. Manosa and A. Saxena (Springer,
Berlin, 2005), p. 177.
\bibitem{igornature} M. van Schilfgaarde, I. A. Abrikosov and B. Johansson, Nature {\bf 400}, 46 (1999).
\bibitem{invar} C. E. Guillaume, Acad. Sci. Paris {\bf 125}, 235 (1897).
\bibitem{invar1} S. Khemlevskyi, A. V. Ruban, Y. Kakehashi, P. Mohn and
B. Johansson, Phys. Rev. B {\bf 72}, 064510 (2005).
\bibitem{invar2} S. Khemlevskyi, I. Turek and P. Mohn, Phys. Rev. Lett.
{\bf 91}, 037201 (2003).
\bibitem{invar3} S. Khemlevskyi, and P. Mohn, Phys. Rev. B {\bf 69},
140404(R) (2004).
\bibitem{sanyal} B. Sanyal and S.K. Bose, Phys. Rev. B {\bf 62}, 12730 (2000); K. Lagarec, D. G. Rancourt, S. K. Bose, B. Sanyal and R. A. Dunlap, J. Magn. Magn. Mater. {\bf 236}, 107 (2001).
\bibitem{noncolin} L. Dubrovinsky, N. Dubrovinskaia, I. A. Abrikosov, M.
Vennstrom, F. Westman, S. Carlson, M. V. Schifgaarde and B. Johnasson,
Phys. Rev. Lett. {\bf 86}, 4851 (2001).
\bibitem{prl} M. L. Winterrose, M. S. Lucas, A. F. Yue, I. Halevy, L. Mauger,
J.A.Munoz, J. Hu, M. Lerche and B. Fultz, Phys. Rev. Lett. {\bf 102},
237202 (2009).
\bibitem{rsg} S. Khmelevskyi and P. Mohn, Phys. Rev. B {\bf 68}, 214412 (2003).
\bibitem{kohn} P. Hohenberg and W. Kohn, Phys. Rev. B {\bf 136}, 864 (1964).
\bibitem{kohn1} W. Kohn and L. J. Sham, Phys. Rev. A {\bf 140}, 1133 (1965).
\bibitem{dlm} B. L. Gyorffy, A. J. Pindor, J. B. Staunton, G. M. Stocks and H.
Winter, J. Phys. F: Met. Phys. {\bf 15}, 1387 (1985); J.B.Staunton,
D. D. Johnson and B. L. Gyorffy, J. Appl. Phys. {\bf 61}, 3693 (1987).
\bibitem{emto1} L. Vitos, Phys. Rev. B {\bf 64}, 014107 (2001).
\bibitem{emto2} L. Vitos, I. A. Abrikosov and B. Johansson, Phys. Rev. Lett.
{\bf 87}, 156401 (2001).
\bibitem{emto3} L. Vitos, H. L. Skriver, B. Johansson and J. Kollar,
Comp. Mat. Sci. {\bf 18}, 24 (2000).
\bibitem{Taylor} D. W. Taylor, Phys. Rev. {\bf{156}}, 1017 (1967).
\bibitem{lapw1} E. Wimmer, H. Krakauer, M. Weinert and A. J. Freeman,
Phys. Rev. B {\bf 24}, 864 (1981).
\bibitem{elk} ELK code, http://elk.sourceforge.net/
\bibitem{lapw2} L. Nordstr\"{o}m and D. J. Singh, Phys. Rev. Lett. {\bf 76},
4420 (1996).
\bibitem{lapw3} L. Nordstr\"{o}m and A. Mavromas, Europhys. Lett. {\bf 49},
775 (2000).
\bibitem{lsda} J. P. Perdew in {\it Electronic structure of solids}, edited
by P. Ziesche and H. Eschrig (Akademic Verlag, Berlin, 1991) pg. 11.
\bibitem{fcd} L. Vitos, {\it Computational Quantum Mechanics for Materials
Engineers}, Springer-Verlag, London, 2007.
\bibitem{mp} M. Methfessel and A. T. Paxton, Phys. Rev. B {\bf 40},  3616 (1989).
\bibitem{weiss} R. J. Weiss, Proc. Phys. Soc. {\bf 82}, 281 (1963).
\bibitem{tc} D. M. K. Paul, P. W. Mitchell and S. A. Higgins, J. Magn. Magn.
Mater. {\bf 54-57}, 1171 (1986).
\bibitem{ebert} S. Polesya, S. Mankovsky, O. Sipr, W. Meindl, C. Strunk and H. Ebert, 
Phys. Rev. B {\bf 82}, 214409 (2010).
\end{thebibliography}
 \end{document}